\documentclass[preprintnumbers,amsmath,amssymb]{revtex4}
\usepackage{graphicx}
\usepackage{amsmath}
\usepackage{epsfig}
\usepackage{amsfonts}
\usepackage{amssymb}

\begin{document}

\title{Nonequilibrium wetting transition in a nonthermal 2D Ising model}


\author{Jef Hooyberghs$^{1,3}$ and Joseph O. Indekeu$^2$}
\affiliation{$^1$VITO, Flemish Institute for Technological Research,
BE-2400 Mol, Belgium
\\$^2$Institute for Theoretical Physics, Katholieke Universiteit Leuven, BE-3001 Leuven, Belgium
\\$^3$Department WNI, Hasselt University, BE-3590 Diepenbeek, Belgium
}.\\

\date{\today}

\begin{abstract}\noindent
 Nonequilibrium wetting transitions are observed in Monte Carlo simulations of a kinetic spin system in the absence of a detailed balance condition with respect to an energy functional. A nonthermal model is proposed starting from a two-dimensional Ising spin lattice at zero temperature with two boundaries subject to opposing surface fields. Local spin excitations are only allowed by absorbing an energy quantum (photon) below a cutoff energy $E_c$. Local spin relaxation takes place by emitting a photon which leaves the lattice. Using Monte Carlo simulation nonequilibrium critical wetting transitions are observed as well as nonequilibrium first-order wetting phenomena, respectively in the absence or presence of absorbing states of the spin system. The transitions are identified from the behavior of the probability distribution of a suitably chosen order parameter that was proven useful for studying wetting in the (thermal) Ising model.
\end{abstract}

\maketitle

\section{Introduction and motivation}
Nonequilibrium wetting has been the subject of profound investigations in the past few decades \cite{Barato,Hinrichsen,Hinrichsen2,Hinrichsen3,Telodagama}. In many of the models studied, the growth, or depinning, of an interface is described relative to a (usually one-dimensional) substrate. Temporal and spatial correlations in the interface are examined and dynamical universality classes are identified. In many cases, the nonequilibrium character of the phenomenon can be related to the breaking of detailed balance of configurational moves. In this paper we are concerned with one of the simplest ways in which detailed balance can be broken, leading to an intrinsically nonequilibrium system. In particular, after a move in configuration space the system can get trapped in certain configurations when the probability for the reversed move is identically zero. In thermal equilibrium, at finite temperature $T$, the reversed move always has a nonzero probability, proportional to the Boltzmann factor, which features the (finite) energy difference of the initial and final configurations. However, if we leave thermal equilibrium, by imposing constraints on the local absorption or emission of energy, detailed balance may be broken. We can go one step further along this line and leave the thermal context altogether by considering a classical system, say, at zero temperature, and providing a nonthermal mechanism for local energy exchange.

To concretize our proposal, consider a (quasi-)two-dimensional lattice spin system at $T=0$ which is exposed to a photon bombardment from some external source. The photon energies $h \nu$ are limited by a cutoff $h \nu_{max} = E_c$. We assume that a spin hit by a photon may absorb an energy $E \leq h\nu$, so that in all cases $E < E_c$. Conversely, a spin may (always) relax by emitting a photon of arbitrary energy and we assume that photon leaves the plane so that the probability for absorption of emitted photons is negligible. The origin of the energy cutoff in this model is quantum mechanical. Although it is not necessary to invoke quantum mechanics explicitly to provide nonthermal energy fluctuations (random-field or random-bond disorder,  electromagnetic fields, mechanical or chemical oscillators being alternative sources), it is a convenient frame-work for obtaining a sharp energy cutoff. In this manner we arrive at a model in which excitations of energy superior to $E_c$ are excluded, which implies that certain configurations can be trapping or ``absorbing".

In the following we develop this model further and investigate how the character of a wetting transition is modified when thermal fluctuations are replaced by constrained nonthermal ones. We do so using Monte Carlo simulation and start within the context of the exactly solved wetting transition of a system in thermal equilibrium. Our paper is structured as follows. In Section II we test our simulation approach on the critical
wetting transition in the two-dimensional Ising model \cite{Abraham 1980}. Section III is devoted to the definition of the nonthermal model, the analysis of the bulk phases and the observation and characterization of nonequilibrium wetting transitions of various nature. Conclusions are drawn in Section IV.


\section{Wetting transition in the 2D thermal Ising model}

\begin{figure}
[ptbh]
\begin{center}
\includegraphics[
height=2.6843cm,
width=10.4493cm
]%
{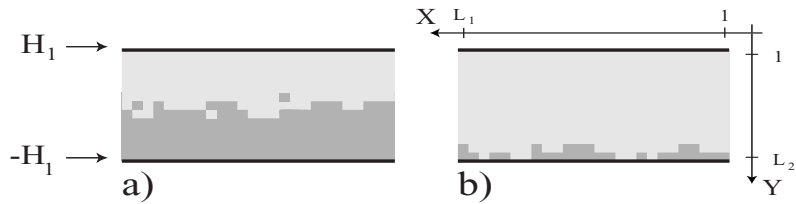}%
\caption{Sketch of typical configurations during Monte Carlo simulations: dark
spots are spin down, light ones are spin up. a) Large surface field $H_{1}$,
complete wetting; b) Small surface field, partial wetting.}%
\label{Fig1}%
\end{center}
\end{figure}

Consider the two-dimensional square lattice Ising model with ferromagnetic nearest-neighbor
interaction $J >0$, at a temperature $T$ below the bulk critical
temperature $T_{c}$ and in zero bulk magnetic field. In this
situation the bulk consists of large coexisting regions of positive
and negative magnetization. In the thermodynamic limit, the
behavior of the bulk is independent of the boundary conditions, but
this is not the case for the interface between the coexisting
phases. In the case of a wetting transition, the surface excess free
energy depends in a singular way on a surface field. To get
efficient computational access to this transition, we use the same setup as in
\cite{albano 1989}: a two-dimensional $L_{1}\times L_{2}$ square lattice of
spins, {\em periodic boundary conditions} along the $X$-axis, {\em open
boundary conditions} along the $Y$-axis, and an {\em anti-symmetric
surface magnetic field} $H_{1} \geq 0$ acting on the spins along the open boundaries (see
Figure \ref{Fig1}). This set-up is often referred to as one with ``opposing boundaries" or ``competing walls"
and possesses surprisingly subtle and rich surface {\em and} bulk cooperative behavior \cite{PE,BLF,SOI,RI,CD}.

In the thermodynamic limit, letting
$L_{1}\rightarrow\infty$ followed by $L_{2}\rightarrow \infty$, a
sharp surface phase transition occurs as a function of the control parameter
$H_{1}$, assuming fixed $T < T_c$. For small and opposing surface fields $H_{1}$ and -$H_1$, with $H_1$ below the wetting point
$H_{1}^{w}$, the interface is localized at one of the boundaries and
this two-fold degenerate state is called partial wetting: Figure \ref{Fig1} b).  For
$H_{1}>H_{1}^{w}$ it is (free-)energetically favorable for the interface
to wander away from the boundaries and the system is in the complete wetting state:
Figure \ref{Fig1} a). For every temperature $T<T_{c}$ such a
wetting point $H_{1}^{w}(T)$ exists for which the excess surface free energy is
singular and for two-dimensional equilibrium systems with short-range interactions this
transition is known to be typically of second order.%

Usually, to characterize accurately a phase transition in a
simulation approach, one studies systems of different (large) sizes
and analyzes the results adopting a finite-size scaling strategy. For our current
reconnaissance study, however, it suffices to verify the order of the
transition and to obtain an estimate of its location in parameter
space. Therefore, a simpler approach is adopted. First, a bulk-like order
parameter, the ``magnetization amplitude" $\Delta$ is introduced
\begin{equation}
\Delta=\frac{\sum_{x=1}^{L_{1}}\left|  \sum_{y=1}^{L_{2}}s(x,y)\right|
}{L_{1}L_{2}}. \label{oderparam delta}%
\end{equation}
where $s(x,y)$ is the spin observable ($=\pm1$) at site ($x,y$). For a
system in the partial wetting state with an interface close to the
boundary, $\Delta$ will be close to its maximum value of $1$, while for
complete wetting with an interface near the middle of the strip, $\Delta$ will be (much) smaller. Next, as a crude
approximation to the successive limits
$L_{1}\rightarrow\infty$ and $L_{2}\rightarrow\infty$ the
fixed values $L_{1}=100$ and $L_{2}=10$ are used, and simulations are
executed for fixed temperature $T$ and different values of $H_{1}$.
During each run a time averaged probability density $P(\Delta)$ of
the order parameter is approximated by a normalized histogram.%

\begin{figure}
[h]
\begin{center}
\includegraphics[
height=3.3486in,
width=4.4823in
]%
{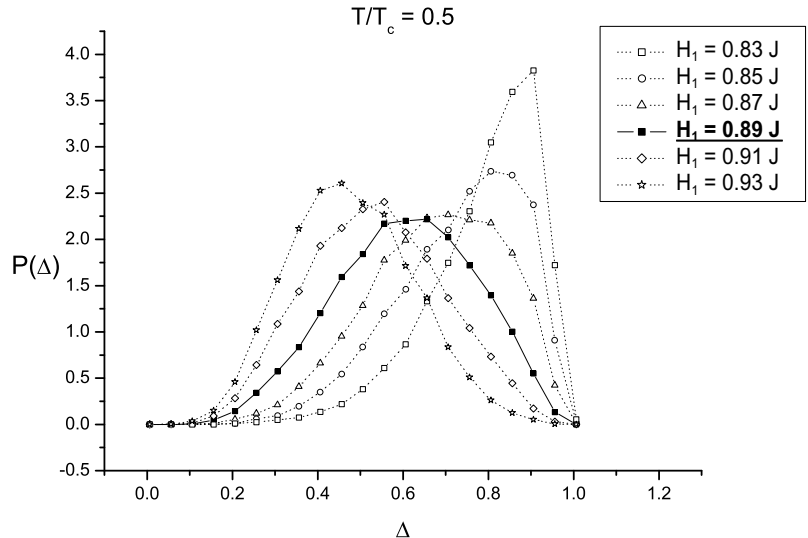}%
\caption{Histograms, or probability distributions, $P(\Delta)$ of the order parameter $\Delta$, for the two-dimensional
Ising model with $L_{1}=100$ and $L_{2}=10$. Each curve is produced from a
simulation of $5.10^5$ Monte Carlo steps/site. The suggested location of the wetting transition is at $H_1 = 0.89 J$, for which the variance of $\Delta$ is maximal.}%
\label{Fig2}%
\end{center}
\end{figure}

In Figure \ref{Fig2} the resulting histograms are shown for $T/T_{c}=0.5$.
For $H_{1}$ significantly smaller than
$0.89 J$ the distributions are centered around a value close to unity,
while for substantially larger surface fields, it is very unlikely to measure a
$\Delta$ close to 1. Around $H_{1}=0.89 J$ one observes a transition
region where the distributions are broad, indicating the presence of
large fluctuations consistently with a second-order interfacial phase transition.
If we identify the wetting transition in the Ising model heuristically with the
point where the distribution of $\Delta$ has a maximal variance, the
result is in quite satisfactory agreement with the exact location. Note that in the limit $T \downarrow 0$ the wetting  transition, at $H_1=J$, is, exceptionally, of {\em first order} and purely determined by minimum energy considerations.
\begin{figure}
[ptbh]
\begin{center}
\includegraphics[
height=4.547cm,
width=5.485cm
]%
{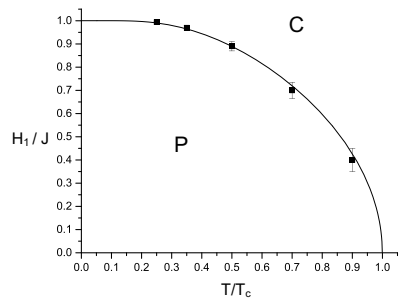}%
\caption{Phase diagram of the wetting transition in the two-dimensional Ising
model, in the variables reduced temperature $T/T_c$ and reduced surface field $H_1/J$, where $J$ is the bulk nearest-neighbor interaction. $ P$: partial wetting phase, $C$: complete wetting phase. The full line is
the exact solution \cite{Abraham 1980}, the squares with error bars are the
estimates from our Monte Carlo simulation and criterion of maximal order parameter variance.}%
\label{Fig3}%
\end{center}
\end{figure}
In Figure \ref{Fig3} our estimates for the transition point for
several temperatures are compared with the exact result of Abraham for the critical wetting phase boundary
\cite{Abraham 1980},
\begin{equation}
e^{2J/kT}= (\cosh(2J/kT)-\cosh(2H_1/kT))\sinh(2J/kT)
\end{equation}
For temperatures not too close to $T_{c}$ we are able to determine
the wetting transition fairly accurately. Closer to the critical
temperature the interface is more fuzzy and it becomes increasingly
difficult to differentiate between partial and complete wetting in a
small system as we used. The method appears to be useful in
principle for locating the wetting transition qualitatively. This is
corroborated by the fact that the shape of the histogram provides an
indication of the {\em order} of the transition. If, like in the
thermal Ising case, one only observes distributions with one maximum
and notices an increase of the variance in between the two sets of
sharper shapes corresponding to the partial wetting ($P$) or
complete wetting ($C$) states, one is in all likelihood dealing with
a second-order phase transition. Close to a first-order transition,
on the other hand, one rather expects distributions with two maxima
that exchange dominance on crossing the transition point
\cite{Wilding}. Further, the valley between the maxima is an
expression of the hysteresis effect, the strength of which is
proportional to the peak-to-valley ratio. Hysteresis is a clear
signature of a first-order phenomenon. We will make use of these
criteria in the analysis of the nonequilibrium model that we now
introduce.


\section{A nonequilibrium wetting transition}

\subsection{Definition of the model}
In our preliminary simulation study of the (thermal) Ising model in the previous section the
statistics was defined through a dynamics described by a
continuous-time Markov process. In such an approach one aims to
replace the canonical-ensemble average of the system by the time
average in the long-time limit. Usually a dynamics is chosen that
locally changes spin configurations $\left\{ s\right\}$ with rates
that obey detailed balance with respect to the Gibbs measure defined
by the energy functional $E(\left\{ s\right\})$. A physically
appealing choice for realizing this is the single-spin flip Glauber
dynamics \cite{glauber}. One chooses at random a spin and calculates
the energy difference $\Delta E$ of the system upon flipping the spin.
Next, time is increased by one unit (usually 1/volume) and the spin
under consideration is flipped with
probability $P_{\text{flip}}$:%

\begin{figure}[h]
\begin{gather}
P_{\text{flip}}=\frac{1}{2}\left[  1+\tanh\left(  -\frac{\beta.\Delta E}%
{2}\right)  \right], \label{flip Glauber}\\%
{\parbox[b]{\textwidth}{\begin{center}
\includegraphics[
height=1.2177in,
width=2.8885in
]%
{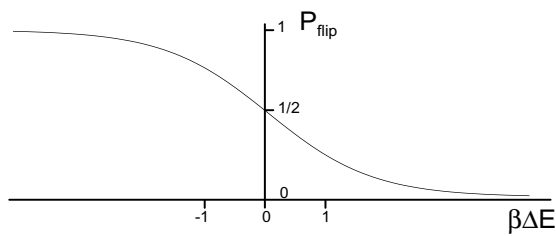}%
\end{center}
\caption{Traditional Glauber spin-flip probability as a function of the reduced local energy difference of final and initial states, for the Ising model in thermal equilibrium. Note that the spin-flip probability is non-zero even for energetically highly unfavorable flips, consistently with the detailed balance property.}%
\label{spinflipT}
\nonumber}}%
\end{gather}
\end{figure}
\noindent where $\beta=1/k T$, with $k$ the Boltzmann constant. On repeating these spin-flip trials,
starting from any initial configuration, in the long-time limit the
system will visit all configurations with the correct Gibbs
probability, i.e., the stationary state (probability distribution) of
the Markov process equals the Gibbs measure.

We now introduce a
different spin-flip dynamics for which the stationary state cannot
be described by a Gibbs measure for any energy functional. In such a
system the dynamics is not induced by thermal effects and it is therefore
an intrinsically nonequilibrium system. It is well known that nonequilibrium
phase transitions can occur \cite{dickman,schutz}, i.e.,
expectation values of observables in the stationary state can depend
in a singular way on control parameters. Our study focuses on a nonequilibrium transition of
wetting type.

From the probability function (\ref{flip Glauber}) it is clear that even if
the energy cost for a spin flip is large, the flip probability remains
strictly positive. This is an essential property of equilibrium models (not
only of the Glauber dynamics) and it is related to the ergodicity of the
dynamics. Therefore, a model without this property can be expected to have qualitatively different behavior. This
leads us to the following definition of a nonequilibrium model.

Consider the
two-dimensional Ising model with nearest-neighbor interaction energy $J$ on the square lattice. We use the
same setup as in Figure \ref{Fig1}, but allow nearest-neighbor spins on
the boundaries $(y=1$ or $y=L_2)$ to have a different nearest-neighbor interaction energy, $J_{\Gamma}$. The energy functional of this system can be written as%
\begin{align}
E\left(  \left\{  s\right\}  \right)   &  =-J\sum_{\left\langle \vec
{r}_{1},\vec{r}_{2}\right\rangle \notin\Gamma}s\left(  \vec{r}_{1}\right)
s\left(  \vec{r}_{2}\right)  -J_{\Gamma}\sum_{\left\langle \vec{r}_{1},\vec
{r}_{2}\right\rangle \in\Gamma}s\left(  \vec{r}_{1}\right)  s\left(
\vec{r}_{2}\right) \nonumber\\
&  -H_{1}\sum_{x}s\left(  x,1\right)  +H_{1}\sum_{x}s\left(  x,L_{2}\right)
\end{align}
where $J$ and $J_{\Gamma}$ are the nearest-neighbor bulk and surface
interaction energies, respectively, and $H_{1}>0$ is the (anti-symmetric) surface field. The two boundaries are denoted by $\Gamma$ and located at $y=1$ and $y=L_{2}$. On this model we define the
dynamics, which is not induced by thermal effects but by local absorption or emission of energy quanta. As in the Glauber model, we allow only a single spin flip at a given time. This flip can be the result of a photon-spin collision (absorption) or a spontaneous photon emission. We assume that the absorption probability decreases with energy, mimicking a certain photon frequency distribution, and becomes zero at a finite energy cutoff $E_c$, which is the maximum photon energy. We further assume that relaxation of a spin towards a lower local energy yields a photon that is emitted out of the lattice plane, so that we ignore spatial correlations between spin flips and avoid possible (secondary) absorption of energies greater than the cutoff.

The single spin-flip probability function we propose is given by
\begin{figure}[h]
\begin{gather}
P_{\text{flip}}=\frac{1}{2}\left[  1+\operatorname{sign}\left(  -\frac{\Delta
E}{E_{c}}\right)  .\min\left(  \left|  \frac{\Delta E}{E_{c}}\right|
,1\right)  \right]  \text{.}
\label{flip non-equil}\\%
{\parbox[b]{\textwidth} {\vspace*{1cm}
\includegraphics[
]%
{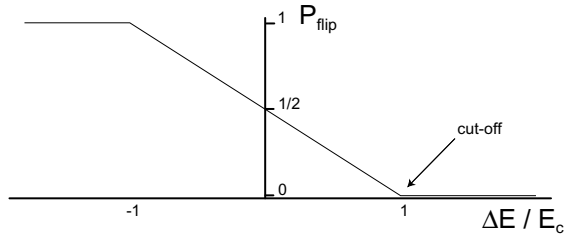}%
\caption{Proposed spin-flip probability as a function of the reduced
local energy difference of final and initial states, for the
nonthermal Ising model. Note that the spin-flip probability is
strictly zero when $\Delta E$ exceeds the energy cutoff $E_c$,
violating the detailed balance principle for thermal equilibrium.}
\nonumber
}}%
\end{gather}
\end{figure}

The important difference between our spin-flip function (Fig.5) and
the traditional one (Fig.4) is not so much in the form of the curve,
which we have chosen piecewise linear just for the sake of
simplicity, but concerns rather the presence of a cutoff energy for
excitations beyond which no spin flips can occur, since $P_{\text{flip}}(\Delta
E/E_c) = 0$ for $\Delta E > E_c$. This has major implications for
the possible transitions between configurations. It can cause the
system to be nonergodic and, as we will show, absorbing states can
occur. Qualitatively, one can ask whether the energy $E_c$ plays a
role similar to that of the energy $2 kT$ in the thermal model, by
examining the spin-flip functions and their slopes at $\Delta E =
0$. On the one hand it seems irrelevant whether we call the energy
scale, relative to which we measure the interactions, $2kT$ or
$E_c$. When both are large compared to the interaction and field
energies the spin-flip function is sampled only over a small domain
near the origin, where its value is about 1/2. In this regime our
model cannot behave differently from the high-temperature limit of
the Ising model. In the opposite limit, when $kT$ and $E_c$ are both
small compared to $J$, $J_{\Gamma}$ and $H_1$, however, the two
models are physically different. In this limit excitations are
excluded in the nonthermal model, while they are not in the
low-temperature regime of the Ising model, although their
probability is small. (The extreme case $T=0$ displays some special
effects, which will be discussed separately further on.) While
minimal energy considerations suffice, in the thermal model, to
discuss phase transitions at low $T$, there is no such thing as an
energy minimization principle in the nonthermal model for small
$E_c$, because the dynamics takes the system to one out of several
absorbing states, regardless of their energy and only depending on
which one is encountered first.

Even greater physical differences between the two models arise when
stable (i.e., ``attracting") non-absorbing states and absorbing
states are {\em cohabitant} (rather than ``coexistent", a term which has a well-defined meaning in the context of equilibrium phases), as for example in the contact process at
sufficiently large infection rate \cite{dickman}. This regime is
found at intermediate values of $E_c$, for which ratios like $J/E_c$
are of order unity. Under these circumstances there is little or no
connection between the behavior of the thermal and nonthermal
models, in addition to the fact that free energy considerations do
not apply to the latter. It is in this regime that our search will
be conducted.

Before we turn to a first analysis of the model, it is convenient to
define dimensionless variables for the couplings and the reduced surface field:
\begin{equation}
K=\frac{J}{E_{c}},\hspace{0.5cm}K_{\Gamma}=\frac{J_{\Gamma}}{E_{c}%
},\hspace{0.5cm}h_{1}=\frac{H_{1}}{E_{c}}\text{.}%
\end{equation}

\subsection{Analysis of the bulk phases}

Since we are interested in wetting transitions, in which an interface can form and move, the bulk of the
system needs to be in the (two-fold degenerate) ordered phase. Therefore we start by
analyzing the bulk behavior of an infinitely large system, defined
by the coupling parameter $K=J/E_c$. It is instructive to recall that the equilibrium critical point of the thermal Ising model is located at $J/kT_c \approx 0.4407 $ (square lattice). We have implicitly reproduced this in our simulations of the critical wetting phase boundary (Fig.3), which terminates at $T=T_c$. In the disordered phase limit, $kT \gg J$ or $E_c \gg J$, the correspondence $E_c \approx 2 kT$ holds, as we discussed. In the spirit of a ``low-$K$" approximation, that is, assuming that the physics found in the high cutoff energy limit can be extrapolated, one would thus expect a dynamical critical point at $K_c \approx 0.22$ for the nonthermal model. However, the appearance of absorbing ordered states above a certain value of $K$ may drastically alter this guess, especially if absorbing states occur already for $K < 0.22$. In this case, ordered bulk phases are being favored by the dynamics and we may expect the dynamical $K_c$ to decrease.

To examine this, in the presence of the cutoff energy $E_{c}$, it
is useful to identify first the type of spin with the highest excitation energy cost $\Delta E$
for flipping: a spin aligned with all its neighbors. If this cost,
$\Delta E=8J$, exceeds $E_{c}$ the spin flip is prohibited.
Consequently, if $K >1/8$ the system has two absorbing states in
which the dynamics is frozen, $\{ \uparrow\}$ and $\{\downarrow\}$: a
configuration with all spins up or all spins down. For $K < 1/8$ there are no absorbing states and we find ourselves in a situation akin to that of the thermal Ising model. Therefore, nonequilibrium effects may be expected to drive the dynamical $K_c$ downwards towards the value 0.125, but not lower.

The next important threshold for $K$ is the value above which a spin
can also not flip if it is aligned with {\em all but one} of its
neighbors: $K > 1/4$. Above this coupling strength infinitely many
($L_2-1$ for a finite system) configurations become absorbing; any
configuration with one horizontal interface separating a spin up
from a spin down region is frozen. This situation is reminiscent of
that of the zero-temperature Ising model, for which the complete
wetting state is ($L_2-1$)-fold degenerate and the interface ``does
not move". Note that for $K < 1/4$ an interface between up and down
domains is not frozen. Consequently, in order to allow interesting
interface dynamics {\em and} the presence of absorbing states, we
turn our attention to the regime $1/8 < K < 1/4$. In this
intermediate-coupling regime the behavior of our model is likely to
be qualitatively different from {\em both} the finite-temperature
and the zero-temperature Ising model.

Now, for consistency, we need to verify that the dynamical critical value $K_c$ above which the system is in an ordered phase is low enough for the regime of bulk order to overlap substantially with the interval $1/8 < K < 1/4$. Since for
$K>1/8$ the ordered states $\{\uparrow\}$ and $\{\downarrow\}$ are
absorbing, it seems that $K_c > 1/8$ is a plausible lower bound. In
the absence of ergodicity it is not a priori clear that a random
initial state at $1/8 \lesssim K $ will evolve into one of the two absorbing states. It is possible
that a third, disordered, stationary state exists which is a dynamical attractor and features an order parameter probability distribution in which the
probability to find the system in an absorbing state is zero.

To
decide on order versus disorder in bulk, we performed a first simulation of the nonthermal model, with uniform couplings $K=K_{\Gamma}$ and $h_1=0$.
We used a square with equal sides of length $L$ and imposed {\em open boundary
conditions} to prevent the system from getting stuck in one of the
states $\{\uparrow\}$ or $\{\downarrow\}$ (for open boundary conditions border spins have only three
neighbors and can still flip against all of them as long as
$K<1/6$). As a simple and efficient order parameter we used the density of broken
bonds
$\rho_{\text{br}}$:%
\begin{equation}
\rho_{\text{br}}=\frac{1}{2L^{2}}\sum_{\left\langle \vec{r}_{1},\vec{r}%
_{2}\right\rangle }\delta\left[  s(\vec{r}_{1}).s(\vec{r}_{2}),-1\right]
\end{equation}
with $\delta$ the Kronecker delta, so that a broken bond is, as usual, defined as a pair of neighboring
anti-aligned spins. In the infinite system limit, $\rho_{\text{br}}$
is zero in the ordered phase and strictly positive in the disordered
one. The scaled
variance of this observable, i.e., multiplied by the number of spins $L^2$, is akin to a dynamical version of the specific heat (capacity) per spin. If the bulk transition is of second order, we expect this dynamical specific heat to diverge approaching the bulk dynamical critical point. For
finite systems we expect a peak, which becomes more
pronounced and moves closer to the correct critical point with increasing
system size.
\begin{figure}
[h]
\begin{center}
\includegraphics[
height=3.2482in,
width=4.2428in
]%
{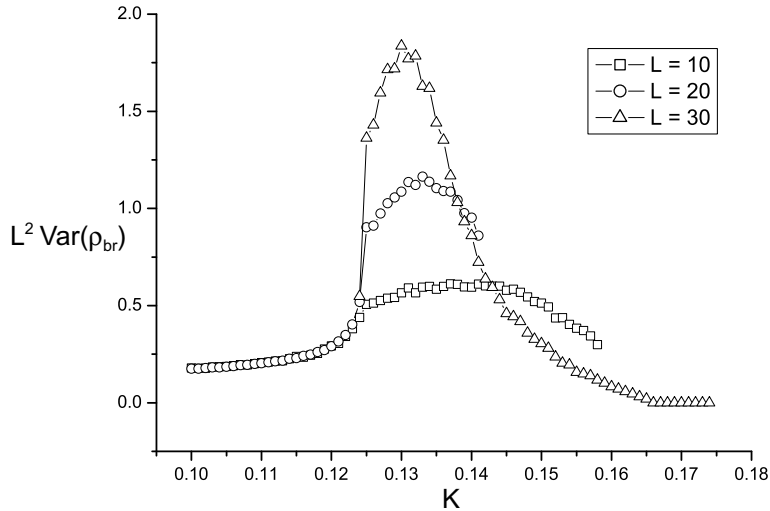}%
\caption{Scaled variance of the density of broken bonds for the nonthermal model on
a square lattice of linear size $L$, as a function of the bulk coupling parameter $K$.}%
\label{Fig4}%
\end{center}
\end{figure}

In Figure \ref{Fig4} the result of the simulation for three system sizes is
shown.\ Besides the weak singularities in $K=1/8=0.125$ and $K=1/6=1.66...$, which reflect effects of the sharp energy cutoff $E_c$,
one clearly observes the building up of a divergence, suggesting a second-order transition with $0.125 \lesssim K_c \lesssim 0.13$. To verify this and to get a precise estimate of the dynamical
critical value $K_c$, larger system sizes and an extrapolation are needed. Both ambitions are outside the scope of this paper. Nevertheless, from
the previous considerations and these data we can convincingly infer that the $K$-window suitable for our explorations is:%
\begin{equation}
0.13\lesssim K < 1/4 \text{.}
\end{equation}
In the remainder of the paper we fix the bulk coupling therefore to a typical value in this range, $K=0.2$, and we now embark on the
investigation of possible wetting phenomena. Note that the spin-flip possibilities for spins on the two boundaries depend on the strengths of surface coupling and reduced surface field, and will be essential ingredients in our study.

\subsection{The wetting transition}

We return to the strip geometry (Fig.1).
In order to understand the effect of the surface interaction energy  $J_{\Gamma} \geq 0$ and
surface field $H_{1} \geq 0$ it is, again, instructive to start with considerations involving the cutoff energy $E_c$. The
energy cost for a surface spin, aligned with all its three neighbors, to flip
is $\Delta E=4J_{\Gamma}+2J-2H_{1}$ or $\Delta E=4J_{\Gamma}%
+2J+2H_{1}$, depending on the spin orientation with respect to the
surface field. If both exceed $E_{c}$, the spins on both
boundaries cannot destroy the absorbing character of the bulk and the
two states $\{\uparrow \}$ and $\{\downarrow\}$ remain absorbing. These
states can be considered as extreme realizations of the {\em partial wetting} state,
hence when
\begin{equation}
4K_{\Gamma}+2K-2h_{1}>1
\label{inequal partial}%
\end{equation}
we concisely say that ``partial wetting is absorbing". For fixed $K$ partial wetting
becomes absorbing if the (reduced) surface field becomes small relative to
the surface coupling. However, this does not imply that partial wetting is also an attractor of the dynamics.

The converse can also take place. Consider a
spin state in which all the spins on the boundaries are aligned and
parallel to the local surface field. This includes, e.g., the state
sketched in Figure \ref{Fig1}a) corresponding to a complete
wetting configuration. We simply refer to all these states as ``complete wetting", regardless of whether the interface is in the middle of the strip or close to a boundary. The important point is that the interface is separated from each boundary by at least one row of spins (cf. the multiply degenerate complete wetting state of the $T=0$ Ising model). Now, when the reduced surface field $h_{1}$ and/or surface coupling $K_{\Gamma}$ becomes
large enough relative to the bulk coupling $K$, the boundary spins
are unable to flip even when the bulk is in the opposite spin state and
the interface can never quite touch the border, but it can still can wander between the boundaries (since $K < 1/4$). The corresponding energy
evaluation gives that ``complete wetting is absorbing" if
\begin{equation}
4K_{\Gamma}-2K+2h_{1}>1.
\label{inequal complete}%
\end{equation}

\begin{figure}
[h]
\begin{center}
\includegraphics[
height=4.5532in, width=6.0272in
]%
{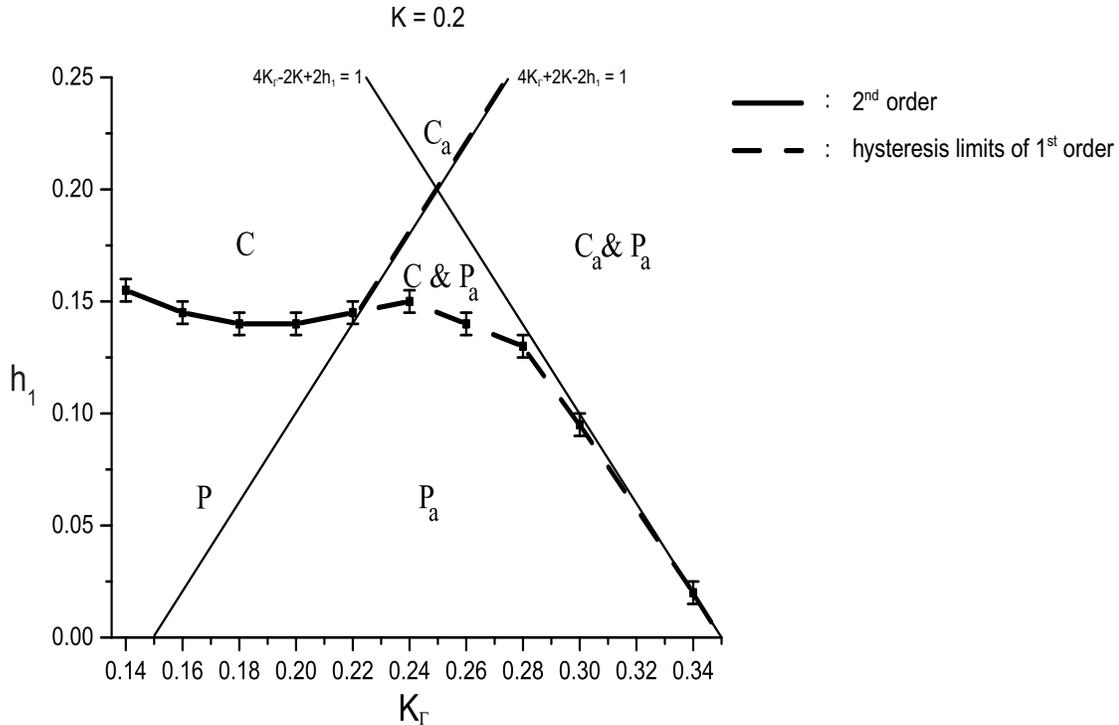}%
\caption{Phase diagram for the wetting transition in the nonequilibrium model
for a fixed bulk coupling $K=0.2$. To the left (for small $K_{\Gamma}$) we encounter a second-order wetting transition from the partial wetting phase ($P$) to the complete
wetting phase ($C$). The thick solid line connecting the ``measured" points is a guide to the eye. For larger $K_{\Gamma}$ the transition changes into a first-order phenomenon, defined by a hysteresis region between two limiting lines (thick dashed lines). In the triangle ($C \& P_a$) complete wetting is an attractor and partial wetting is absorbing, which leads to ``cohabitation" of dynamical states. Another cohabitation is found for larger $K_{\Gamma}$, still in the hysteresis region, where both complete wetting and partial wetting are absorbing states ($C_a \& P_a$). The thin solid lines denote the exact location of the limits of occurrence of absorbing states.}%
\label{Fig5}%
\end{center}
\end{figure}

For $K=0.2$ the regions of absorbing complete and partial
wetting are drawn in Figure \ref{Fig5}. In the area of large surface
coupling and/or (reduced) surface field, there is an ambivalent ``phase" in which both partial and
complete wetting are absorbing states.\ If the system is in either
of these states it can never get out again. The dynamics in this phase is very different from that in an equilibrium system and it is not possible to identify partial or complete wetting phases. A certain fraction of initial configurations evolve into the former and the remainder into the latter. To attempt to identify ``stable" phases based on minimal energy considerations is totally meaningless in this dynamical system \cite{meaningless}.
To construct a phase diagram for the wetting transition we performed
simulations for $K=0.2$ and analyzed them using the histogram
method described above for the Ising model.\ For fixed $K_{\Gamma}$
we made scans along $h_{1}$.
\begin{figure}
[hh]
\begin{center}
\includegraphics[
height=3.4022in,
width=4.2428in
]%
{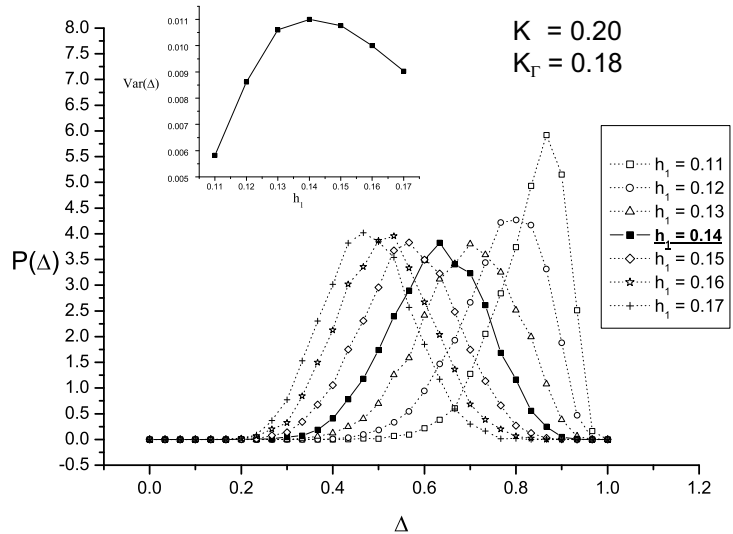}%
\caption{Histograms of the order parameter $\Delta$, for the
nonequilibrium model with $L_{1}=100$ and $L_{2}=10$. Each curve is
produced from a simulation of $2.10^6$ Monte Carlo steps/site.
Inset: variance of $\Delta$ for the different
values of the reduced surface field $h_{1}$.}%
\label{Fig6: k0.18}%
\end{center}
\end{figure}
In Figure \ref{Fig6: k0.18} the result of such a scan is shown for $K_{\Gamma
}=0.18$. Qualitatively, the behavior is similar to that of the thermal Ising model at critical wetting,
i.e., the transition from partial to complete wetting upon increasing $h_{1}$ appears to be of second order.\ In the inset of the plot, one can see that the variance of
the order parameter $\Delta$ (defined in (\ref{oderparam delta})) has a
maximum around $h_{1}=0.14$, which we identify with the transition
point. The same type of transition was found for arbitrary values of $K_{\Gamma}$ below
approximately $0.22$.

For $K_{\Gamma}$ above $0.22$ the wetting transition enters the area of absorbing
partial wetting and its character is drastically changed. There is now an entire subspace
in parameter space characterized by the existence of two {\em cohabitant states} (see Fig.~\ref{Fig5}). In this region complete wetting can be an attractor of the dynamics while partial wetting is absorbing. Let us examine this now in more detail.

Suppose the system is initially in the complete
wetting state in the region labeled $C \& P_a$ in Fig.7. We find that it remains in the complete wetting state for a (very) long time. Now $h_{1}$ is slowly decreased. Below some value of $h_1$ the complete wetting state {\em rapidly} evolves into
the partial wetting state. This happens in the region labeled $P_a$ in Fig.7. Since partial wetting is absorbing
the system is trapped in one of the absorbing states
$\{\uparrow\}$ or $\{\downarrow\}$. In this {\em transition} a histogram of the
order parameter $\Delta$ abruptly changes from a distribution with a
positive variance to a Dirac delta distribution on $\Delta=1$.

The impression that the transition we just discussed is part of a dynamical {\em first-order} phenomenon is corroborated by
the presence of a hysteresis effect: if we now reverse our path and increase
$h_{1}$ so that we re-enter the region labeled $C\& P_a$ in Fig.7, the system stays in the
absorbing partial wetting state. Upon further increase of $h_1$, when crossing the boundary $4K_{\Gamma}+2K-2h_{1}=1$ between the regions labeled $C \& P_a$ and $C$ in Fig.7, it evolves abruptly into the
complete wetting state. Also this reverse transition is part of the dynamical {\em first-order} phenomenon.

When one studies, in a thermal system, a first-order transition in a simulation, one
can use the histogram approach. Around the transition point the
system can be in two stable states. For a finite system, this means
it will stay for some time in one of these states but every now and then
it will jump into the other. In this way, one long simulation can
probe the whole configuration space and one can construct a
histogram that in the neighborhood of the transition will show two
peaks representing the two stable states. On setting the control
parameters further away from the transition values, one of the peaks
will start dominating until the second peak disappears: this sets
the borders of the hysteresis region.
\begin{figure}
[h]
\begin{center}
\includegraphics[scale=0.5
]%
{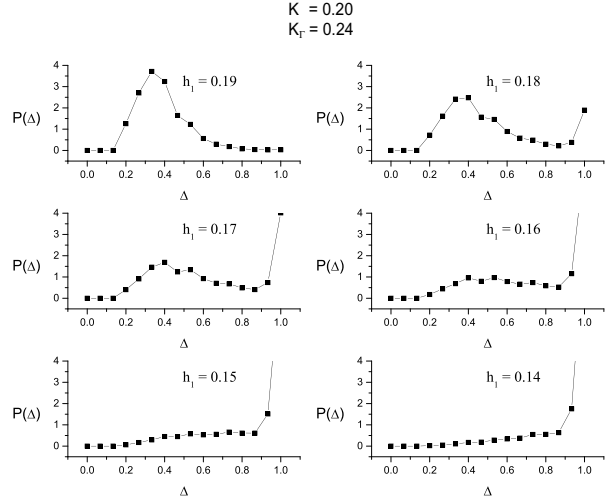}%
\caption{Histograms of the order parameter $\Delta$, for the non-equilibrium
model with $L_{1}=20,L_{2}=5$ and random spin flip probability $P_{r}=0.001$.
Every curve is produced from a simulation of $2.10^6$ Monte Carlo steps/site.}%
\label{Fig 7 k0.24}%
\end{center}
\end{figure}
\begin{figure}
[hh]
\begin{center}
\includegraphics[scale=0.7
]%
{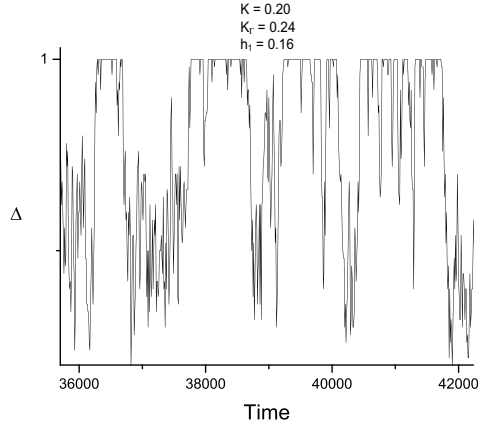}%
\caption{The order parameter $\Delta$ as a function of time for the model with
$h_{1}=0.16$, corresponding to one of the cases (middle right) for which $P(\Delta)$ is shown in Fig.~\ref{Fig 7 k0.24}.}%
\label{Fig8 time}%
\end{center}
\end{figure}

However, our nonthermal system is equipped with absorbing states from
which the spin configuration cannot escape, even for small system sizes. Under these circumstances the
simulation cannot probe the configuration space properly and a
naive use of the histogram method fails. To overcome this technical problem
we introduced random spin flips: with a very low probability $P_{r}$
we flip spins, independently of the energy cost. We have to be
careful with this, since it weakens the action of the cutoff
energy and might restore the second-order character of the wetting transition.
If, however, for very small $P_{r}$ we find non-second order
behavior, we interpret this as evidence that the action of the cutoff is physically responsible for
it and that a first-order dynamical phenomenon takes place.

In our simulation we used $P_{r}=0.001$ and a system of small size, with
$L_{1}=20$ and $L_{2}=5$. (The transition time between two cohabitant
states grows very rapidly with system size and with $1/P_{r}$.) The
results for $K_{\Gamma}=0.24$ are presented in Figure \ref{Fig 7
k0.24}. For $h_{1}>0.18$, in the region labeled $C$ in Fig.7, we found a single-peaked histogram,
representing a complete wetting state. From the moment one enters
the absorbing partial wetting region labeled $C\& P_a$, at $h_{1}=0.18$, a Dirac delta
peak at $\Delta=1$ appears next to the peak of complete wetting. On
decreasing $h_{1}$ this second maximum
eats more and more of the probability distribution until at $h_{1}%
\approx 0.15$, when entering the region labeled $P_a$, the peak of complete wetting disappears. The values
$h_{1}=0.18$ and $h_{1}\approx 0.15$ are the two limits of the first-order
transition phenomenon at the given value of $K_{\Gamma}$ and mark the boundaries of the dynamical hysteresis region. These boundaries are indicated
in Figure \ref{Fig5} by the dashed lines. We stress that the first-order phenomenon we observe pertains to an entire hysteretic region in the phase diagram, delimited by the two abrupt transitions we discussed. In our opinion, it is meaningless to attempt to locate a first-order transition ``line" in the phase diagram, like one is used to do for equilibrium phase transitions. Based on our understanding of the dynamical behavior so far, we believe that a line of this sort does not and should not exist in our model.

Figure \ref{Fig8 time} shows, for a system with parameters inside the part of the hysteresis
region labeled $C\& P_a$, the time evolution of the control parameter from which the histograms are
constructed. One can clearly distinguish the two cohabitant states between which
the systems jumps. Note that complete wetting is an attractor of the dynamics in the sense that even configurations close to the absorbing $\{\uparrow\}$ or $\{\downarrow\}$ states can evolve towards configurations with an interface in the middle of the strip. On the other hand, either one of the (quasi-)absorbing states is also frequently visited. Due to the non-vanishing random spin flip probability $P_{r}=0.001$, these visits are not permanent.

Finally we remark that the separatrices in the phase diagram of Figure
\ref{Fig5} between phases $P$ and $P_{a}$, or between phases $C$ and $C_{a}$, are due
to the singular behavior of the spin flip probabilities (\ref{flip non-equil}) about the value $\Delta E = E_c$ and
are irrelevant to the wetting phenomena. For instance, the transition from $P$ to $P_a$ is a one-dimensional surface transition (which is not a sharp  phase transition).

In closing this section we also show, for completeness and for comparison, the phase diagram of the first-order wetting phase transition in the zero-temperature Ising model, in Figure 11. The equilibrium transition at $H_1=J$ is shown, together with the spinodal lines at $\pm 2J_{\Gamma}/J + 1 = H_1/J$. With respect to the Glauber dynamics all states with an interface parallel to the boundaries and the fully ordered states are absorbing states in the region $P_a \& C_a$ in between the spinodals. Beyond these spinodals, only {\em either} the fully ordered states $\{\uparrow\}$ and $\{\downarrow\}$ ($P_a$) {\em or} the interface states are absorbing ($C_a$).
\begin{figure}
[h]
\begin{center}
\includegraphics[scale=0.5
]%
{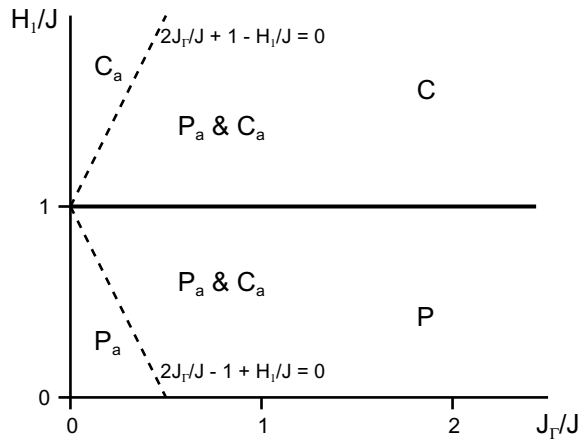}%
\caption{(Trivial) wetting phase diagram of the zero-temperature Ising model, in the plane of surface to bulk coupling ratio $J_{\Gamma}/J$ and surface field to bulk interaction energy ratio $H_1/J$. Partial ($P$; fully ordered, 2-fold degenerate) and complete wetting ($C$; straight interface parallel to the boundaries, $(L_2-1)$-fold degenerate) phases are separated by a first-order phase transition at $H_1=J$. The partial (complete) wetting state is metastable up (down) till the upper (lower) spinodal (dashed line). $P_a$ and $C_a$ denote absorbing phases with respect to the single spin flip Glauber dynamics, which in this zero-temperature limit only allows single spin flips that lower the energy.}%
\end{center}
\end{figure}


\section{Conclusions}
In this paper we studied the wetting transition from partial wetting
to complete wetting in a nonthermal two-dimensional ferromagnetic
Ising-like lattice model defined by a stochastic dynamics. The
geometry is a periodic rectangular strip with opposing walls. We
used a simple Monte Carlo simulation technique which could reproduce
the order of the wetting transition and approximate the known phase
diagram for the case of critical wetting in the equilibrium 2D Ising
model with nearest-neighbor coupling $J/kT$ and reduced surface
fields $\pm H_1/kT$. For this equilibrium test case we employed a
single spin-flip dynamics which obeys detailed balance with respect
to the Gibbs measure for the given energy functional. Next we
introduced an energy functional with, in addition to $J$ and $H_1$,
a tunable surface spin interaction energy $J_{\Gamma}$ (between
nearest neighbors) and introduced a non-equilibrium dynamics based
on a piece-wise linear single spin-flip probability function which
features an energy cutoff for excitations, $E_c$, beyond which no
spin flips can occur. Spin flips concur with photon absorption or
emission, and $kT$ is negligible compared to all other interaction
and field energies involved.

Analyzing the wetting transition using the Monte Carlo simulation
technique, we found a rich phase diagram in the space $(H_1/E_c,
J_{\Gamma}/E_c)$, for fixed nearest-neighbor coupling strength in
the intermediate regime $1/8 < K = J/E_c < 1/4$. In this range of
$K$ the interface between oppositely magnetized domains can still
wander freely, but single spin excitations in a uniform domain are
suppressed by the cutoff. Absorbing states can occur, and partial
wetting as well as complete wetting can be absorbing. A second-order
wetting transition is found for weak surface coupling, when
absorbing states are absent. This transition changes to a dynamical
first-order phenomenon, with hysteresis, for strong surface
coupling. In a special region of the phase diagram, a dynamical
attractor corresponding to complete wetting is {\em cohabitant} with
an absorbing partial wetting state. The first-order phenomenon is
qualitatively new and its properties are largely due to the presence
of the energy cutoff for excitations. Only some of its
characteristics can be understood as remnants of the trivial
first-order wetting transition in the zero-temperature Ising model.
The cartoon shown in Fig.12 summarizes the most important features
of our wetting phase diagram.
\begin{figure}
[h]
\begin{center}
\includegraphics[scale=0.5
]%
{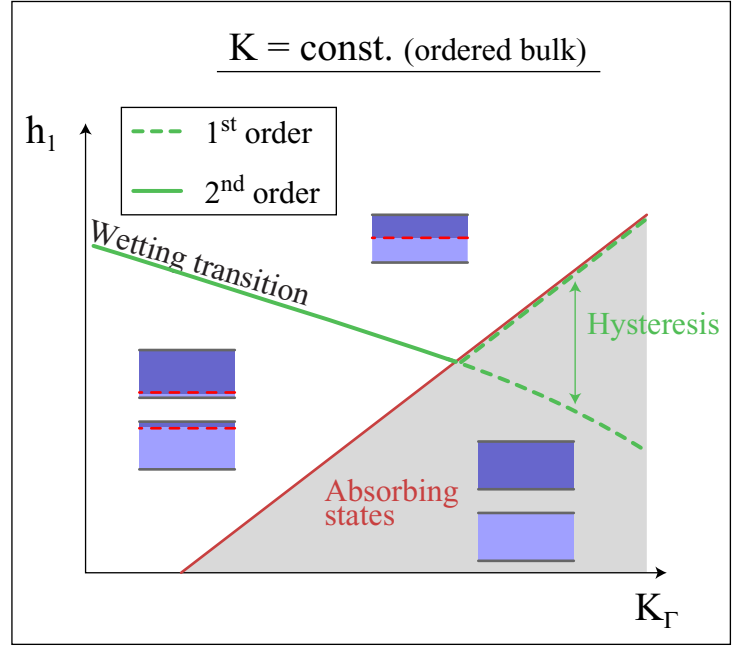}%
\caption{Tutorial sketch of (part of) the wetting phase diagram of the nonthermal 2D Ising model, at fixed (intermediate) bulk coupling $K= J/E_c$, as a function of surface coupling $K_{\Gamma}= J_{\Gamma}/E_c$ and surface field $h_1=H_1/E_c$. From the point of view of wetting phenomena there are three qualitatively distinct regions. A partial wetting phase is found for small $h_1$, a complete wetting phase occurs at large $h_1$ and a hysteresis region appears for large $K_{\Gamma}$. The wetting transition line is of second order, except in the hysteresis region, which is bounded by two lines of abrupt dynamical transitions and which represents a dynamical first-order phenomenon. In the shaded region (in grey) the two extreme partial  wetting states are absorbing.}%
\end{center}
\end{figure}


\newpage

\begin{thebibliography}{99}
\bibitem{Barato} For a recent review of nonequilibrium wetting, see A. C. Barato, J. Stat. Phys. {\bf 138}, 728 (2010); for a tutorial review of nonequilibrium phase transitions see H. Hinrichsen, Physica A {\bf 369}, 1 (2006).
\bibitem{Hinrichsen} H. Hinrichsen, R. Livi, D. Mukamel and A. Politi, Phys. Rev. Lett. {\bf 79}, 2710 (1997); Phys. Rev. E {\bf 61}, R1032 (2000); Phys. Rev. E {\bf 68}, 041606 (2003).
\bibitem{Hinrichsen2} C. Gogolin, C. Meltzer, M. Willers and H. Hinrichsen, Phys. Rev. E {\bf 79}, 041111 (2009).
\bibitem{Hinrichsen3} A. C. Barato, H. Hinrichsen and M. J. de Oliveira, Phys. Rev. E {\bf 77}, 011101 (2008).
\bibitem{Telodagama} F. de los Santos, M. M. Telo da Gama and M. A. Mu$\tilde{\rm n}$oz, Phys. Rev. E {\bf 67}, 021607 (2003).
\bibitem{Abraham 1980}D.B. Abraham, Phys. Rev. Lett. \textbf{44} (1980) 1165.
\bibitem{albano 1989}E.V. Albano, K. Binder, D. W. Heermann and W. Paul,
Surface Science \textbf{223}, 151 (1989).
\bibitem{PE} A.O. Parry and R. Evans, Phys. Rev. Lett. {\bf 64}, 439 (1990); Physica A {\bf 181}, 250 (1992).
\bibitem{BLF} K. Binder, D. P. Landau and A. M. Ferrenberg, Phys. Rev. E {\bf 51}, 2823 (1995).
\bibitem{SOI} M.R. Swift, A. Owczarek, and J.O. Indekeu, Europhys. Lett. {\bf 14}, 475 (1991).
\bibitem{RI} J. Rogiers and J.O. Indekeu, Europhys. Lett. {\bf 24}, 21 (1993).
\bibitem{CD} E. Carlon and A. Drzewinski, Phys. Rev. Lett. {\bf 79},1591 (1997).
\bibitem{Wilding} N.B. Wilding, Am. J. Phys. {\bf 69}, 1147 (2001).
\bibitem{glauber}R. Glauber, J. Math. Phys. \textbf{4}, 294 (1963).
\bibitem{dickman} J. Marro and R. Dickman, ``Nonequilibrium Phase Transitions
in Lattice Models", Cambridge University Press, Cambridge (1999).
\bibitem{schutz}G. M. Sch\"{u}tz, in ``Phase Transitions and Critical
Phenomena", Vol.19, edited by C. Domb and J. L. Lebowitz, Academic, Londen (2000).
\bibitem{meaningless} Incidentally, the absorbing partial wetting state and the lowest energy complete wetting configuration minimize the total spin energy $E(\{s\})$ for $h_1 < K$ and  $h_1>K$, respectively.
\end{thebibliography}
\end{document}